# Photonic Neuromorphic Accelerators for Event-Based Imaging Flow Cytometry

I. Tsilikas[1,2,‡], A. Tsirigotis[1‡], G. Sarantoglou, S. Deligiannidis[3], A. Bogris[3], C. Posch[4], G. Van den Branden[4], C. Mesaritakis[1]*

[1] Department of Information and Communication Systems Engineering, University of the Aegean, Palama 2, 83100, Samos - Greece  [2] Department of Physics, School of applied mathematical and physical sciences, Zografou Campus, 157 80, Athens - Greece
[3] Department of Informatics and Computer Engineering, University of West Attica, Ag. Spyridonos, Egaleo – Greece
[4] Prophesee Metavision, rue du faubourg Saint-Antoine 74, 75012 Paris-France

[*] cmesar@aegean.gr
[‡] these authors contributed equally to this work

## Abstract

In this work, we present experimental results of a high-speed label-free imaging cytometry system that seamlessly merges the high-capturing rate and data sparsity of an event-based CMOS camera with lightweight photonic neuromorphic processing. This combination offers high classification accuracy and a massive reduction in the number of trainable parameters of the digital machine-learning back-end. The event-based camera is capable of capturing 1 Gevents/sec, where events correspond to pixel contrast changes, similar to the retina's ganglion cell function. The photonic neuromorphic accelerator is based on a hardware-friendly passive optical spectrum slicing technique that is able to extract meaningful features from the generated spike-trains using a purely analogue version of the convolutional operation. The experimental scenario comprises the discrimination of artificial polymethyl methacrylate calibrated beads, having different diameters, flowing at a mean speed of 0.01m/sec. Classification accuracy, using only lightweight, FPGA-compatible digital machine-learning schemes has topped at 98.2%. On the other hand, by experimentally pre-processing the raw spike data through the proposed photonic neuromorphic spectrum slicer at a rate of $3\times10^6$ images per second, we achieved an accuracy of 98.6%. This performance was accompanied by a reduction in the number of trainable parameters at the classification back-end by a factor ranging from 8 to 22, depending on the configuration of the digital neural network. These results confirm that neuromorphic sensing and neuromorphic computing can be efficiently merged to a unified bio-inspired system, offering a holistic enhancement in emerging bio-imaging applications.

## Introduction

Imaging flow cytometry (IFC) is the physical evolution of conventional flow cytometry (FC)[1] that strengthens the light-scatter recording capabilities of FC with accurate recording of the morphological features of the particles under investigation. IFC significantly enhances the detection capabilities of typical cytometers[2]. Furthermore, visualization of spatial information adds an additional layer of qualitative information and allows the simultaneous recording of both brightfield and darkfield images[3,4] without hindering typical fluorescent-based analysis. Furthermore, the high throughput of IFC schemes minimizes the lab-to-diagnose time to few minutes, unlocks the use of ultra-low sample volumes, whereas more importantly allows the detection of very rare cell populations[5], such as cancer cells in blood flow[6], profile complex phenotypes or capture the associated dynamic of cell development phases[1,7]. Exploiting these merits, IFC has started infiltrating a broad range of applications among which the most prominent are new drug discovery[8], personalized medicine, DNA sequencing and rapid disease diagnostics[3,9,10]. In principle, the above applications can be partially addressed by conventional high-throughput FC, but in most cases, a fluorescent agent is required that in turn can affect the molecule/cell's chemical or biological properties and thus can compromise the detection sensitivity or render it difficult to be applied. Furthermore, conventional FCs are mechanically complex and costly, demand high sample volume due to the fact that their output is based on statistics, mandate the use of elaborate microfluidic systems and last but not least can be operated only from trained personnel. This last aspect is of paramount importance, as easy-to-handle home-care diagnostics are a rapidly growing necessity in modern societies with an aging population.

As a response, during the last years, a growing number of IFC modalities have emerged, that try to simultaneously address two contradicting requirements: namely high-speed operation and crisp spatial feature capturing. The contradiction is based on the fact that conventional capturing devices, experience blur effects, when high frame rate is requested, due to their limited bandwidth and high latency. Aiming to amend this, a wide pallet of technologies has been proposed, each offering vastly different performance metrics, in terms of the maximum particle flow that they can handle, footprint and wavelength range that they can operate. Broadly, IFC technologies can be categorized into two main categories: Single detector/pixel IFCs and 2-dimensional (2D) IFCs. The first category is dominated by time-stretched based approaches, where spatial information is projected in time and it is detected through a single high-speed photodiode. In this case, spatial resolution is linked to temporal resolution and thus it can proliferate by the high-speed capabilities of photonic technology[11,12], theoretically offering detection of particle flow $>10^6$ particles/s. Following a similar concept, dual comb microscopy schemes, projects spatial information to the consecutive radio-frequency (RF) beating of two optical combs with different free spectral range[13,14]. Despite the discrete merits of both techniques, they both mandate complex optical setups, costly optical sources, are restricted to the near-infrared (NIR) wavelengths, while both, in practice, generate massive amount of data during acquisition. In the 2D IFC category, time delay and integration (TDI) cameras allow motion-blur free images but at the cost of low particle-speed (<3000 particle/s), this trade off stems from the necessity of high-integration time to unlock sufficient gain at the detector[15]. Alternatively, schemes based on strobe-photography can unlock high rates exceeding 50000 particles/s but at the cost of using sophisticated microfluidic schemes so as to enable precise control of the particle motion (trajectory and speed)[16]. Similarly, to the 1D case, 2D modalities also in typical cases generate an increased volume of data.

Recently, an alternative type of camera has been proposed for IFC, which follows a bio-inspired acquisition principle that relies on autonomous and asynchronously recording contrast detection events, instead of the camera periodically transmitting every pixel's intensity in the form of a frame[17]. In detail, event-based cameras (neuromorphic cameras) detect the contrast variation among adjustment pixels and transmit, in an asynchronous manner electrical spikes, similar to the retina-ganglion cells in mammals. Through this technique, time-continuous data transmission is limited to pixels detecting temporal contrast events in their field of view, thus massively reducing transmission bandwidth requirements, while offering wide dynamic range acquisition and radically increased capturing rate (i.e. temporal resolution) to IFC compatible levels [2,18,19]. More importantly, these types of cameras rely on standard CMOS technology, thus are relative low-cost, have limited footprint and do not require complex optical systems or sophisticated control electronics. A critical difference in this IFC modality is that bio-inspired sparsity (spike encoding) can potentially reduce the volume of data generated during measurement.

Another critical step in IFC involves the data processing techniques employed after data capture. In this context, requirements vary depending on the specific IFC installation. For example, generic microfluidics permit multiple particles in the field of view, necessitating the use of object tracking algorithms[20]. In addition, the acquisition speed and resolution of the generated signal (frame rate and pixel count), impose additional restriction on the size of the following neural network and on the associated processing latency. A plethora of neural network architectures have been proposed so as to efficiently tackle the aforementioned requirements, ranging from convolutional neural networks (CNNs) to recurrent neural networks (RNNs) and deep neural networks (DNNs)[21,22]. Each of these neural network implementations is adapted to the IFC, offering a diverse mix of advantages and disadvantages in terms of complexity, accuracy, and latency. However, features such as a limited number of floating-point operations (FLOPs), low latency, low power consumption and FPGA compatibility are considered critical for efficient, real-time and handheld IFCs. In all cases, an increased data volume generated by the IFC modality leads to a higher parameter count in the neural network, which in turn results in unavoidably higher power consumption and latency during neural network training.

In this work, we realize a low-cost, compact, light emitting diode (LED) based IFC scheme built around a CMOS neuromorphic camera, capable of capturing up to 1Gevents/sec with μs-range temporal resolution, delivering an equivalent of 100 kframes/sec[17]. In this case, the IFC signals correspond to asynchronous electrical spikes, signifying individual pixel's relative intensity changes (aka temporal contrast). The developed IFC module utilizes a generic off-the-shelf microfluidic chip with a single straight channel, without any sheath flow control mechanism. The IFC scenario realized was discriminating aqua solutions of Polymethyl methacrylate (PMMA) spheres of 12, 16 and 20μm in diameter. The generated electrical spikes (events) were processed in two distinct ways: the first method involved direct processing using lightweight machine-learning (ML) models, such as simple fully connected layers (FCLs) and compact RNNs. This direct method yielded high classification accuracy, up to 97.6% and 98.6% for the optimal FCL and RNN configurations, respectively. However, direct processing with FCLs required 30K trainable parameters, and RNNs required 1M trainable parameters, alleviating the hardware requirements and power consumption during training. The second approach adopts an unconventional path by transferring the IFC signals into the optical domain, where they are pre-processed through a photonic neuromorphic scheme utilizing an optical spectrum slicing (OSS) architecture[23]. The pre-processed outputs are then also fed into a conventional digital FCL as above. In this scenario, the photonic neuromorphic pre-processor acts as an analogue CNN accelerator, achieving an even higher classification accuracy of 98.6% compared to standalone FCL. Additionally, this approach resulted in a significant reduction in the number of trainable parameters in the FCL by a factor of >20. These experimental results validate that the combination of neuromorphic sensing and processing can enhance accuracy and more importantly generate a massive impact on the power consumption requirements of the overall IFC schemes.

## Results

**Neuromorphic Camera based IFC**

The experimental IFC setup is depicted in Fig. 1a-b. The light source is a simple LED emitting at 635 nm, chosen to simplify the experimental system while avoiding the generation of laser-source, diffraction patterns that could obscure the actual physical features of the particles. Two microscope objectives were utilized to focus and collect light into/from a generic microfluidic channel. The three classes of particles tested were PMMA spheres with diameters of 12, 16, and 20 μm (refer to the Methods section for details). A steady flow was maintained in the channel with the aid of a vacuum pump, allowing the particles to reach mean velocities ranging from 1 to 0.07 m/s, corresponding to an ideal particle flow rate of 500 to 350 particles/s (assuming an ideal sequential flow of particles with no gaps between them). However, the actual measured flow rate was limited to 10-15 particles/s, a limitation necessitated by the need to significantly dilute the solutions to prevent clogging. The event-based camera (Prophesee Gen4 sensor) provided a spatial resolution of 640 x 480 pixels and a temporal resolution of 1 μs per pixel, resulting in an equivalent capturing rate of 1 Gevents/sec[17].

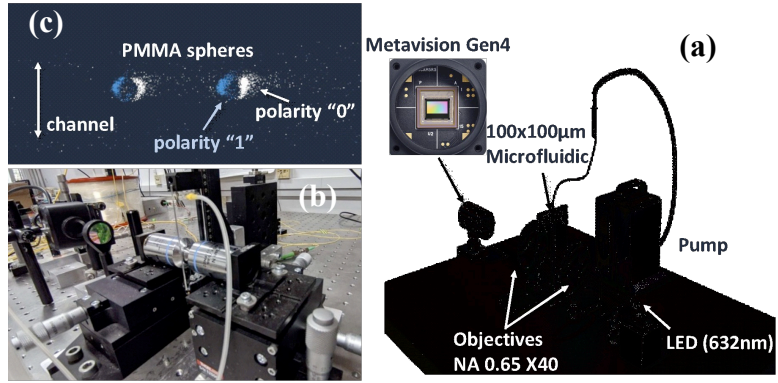

Figure 1. (a) Schematic of the proposed IFC: the neuromorphic camera used is a Prophesee Gen4 sensor[17]. (b) The installed IFC. (c) Typical recorded synthetic frame depicting 20μm PMMA spheres flowing within the microfluidic channel: polarity "1" and "0" corresponds to pixel contrast increase and decrease respectively.

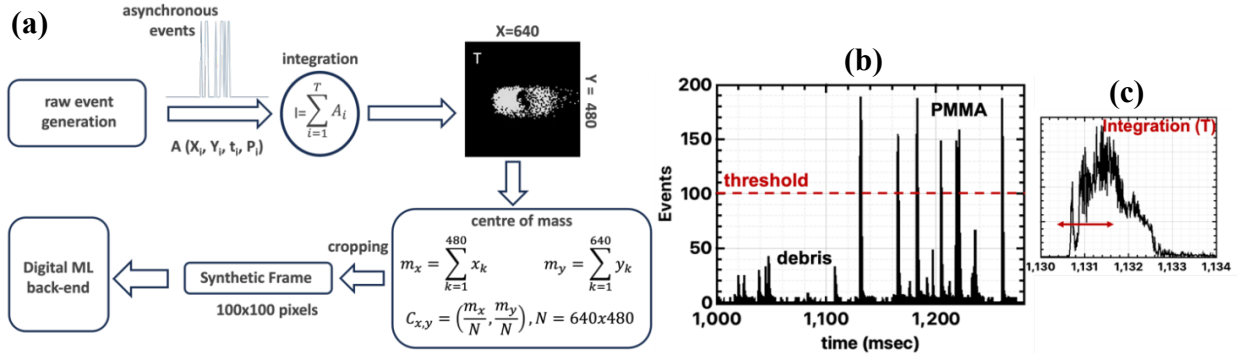

Figure 2. (a) Data collection and lightweight pre-processing pipeline. (b) Total number of spikes across all 2D coordinates over time for a typical sample of 20 μm PMMA spheres. (c) Zoomed-in version of a single peak from part b.

The lightweight data processing pipeline is illustrated in Figure 2. The Prophesee camera generates asynchronous events – tuples that include the coordinates, a timestamp, and a binary polarity signifying whether the pixel contrast is triggered by an intensity increase or decrease (X, Y, t, P). In our case, we sum all events per pixel (regardless of their polarity), over an integration time (T) to generate I(X,Y). The choice of T depends on the speed of the particles and is linked to the number of events per frame; meaning that for a fast-moving sphere, a low value of T would result in too few events per synthetic frame, thus a lower signal-to-noise ratio (SNR). Conversely, a high value of T obscures the recording of the fine spatial features of the particles, similar to motion blur. To compute the optimum T for our setup, we recorded the total number of spiking events over time (see Fig. 2b). This measurement is performed once, and the particles' flow is represented by individual peaks over time. Fig. 2b facilitates the extraction of two key observations.

The first concerns the duration of each peak, which remains consistent across all recorded peaks, regardless of their amplitude, and corresponds to the time needed for each object to enter and exit the IFC's field of view (Fig. 2c). Therefore, the rise time of these peaks can provide an estimate of the optimum integration window (T) for the specific vacuum pump setting; in our case, T was set to 3 ms (Fig. 2c). The second observation is that the peaks' amplitude (number of events) varies, with several peaks observed in Fig. 2b having a significantly low number of events (<50). These peaks do not correspond to actual PMMA particles but are debris or air bubbles contaminating the PMMA mixture, evident in all samples. Therefore, a unified criterion of rejecting events with a low event count (<100) has been used throughout all measurements.

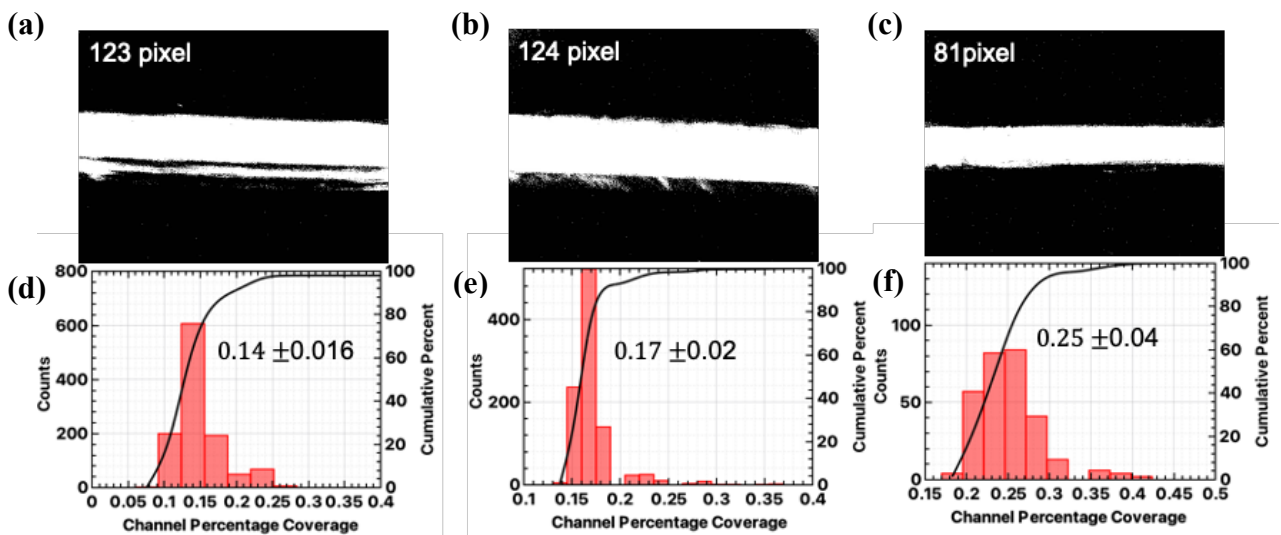

Figure 3: a-c: Images of microfluidic channels from different experimental instances, showing variations in width (in pixels) due to minor misalignments. d-f: Histograms of the distribution of particle widths, normalized to the width of the corresponding channel for each particle class: d) PMMA spheres of 12 μm diameter, e) 16 μm, and f) 20 μm.

Interestingly, the information in $I$ follows a rate-encoding scheme, where pixels with a high spike count correspond to spatial locations subjected to strong activity (Fig. 2a). Furthermore, considering we did not choose a sophisticated microfluidic channel; the particles do not flow in a single file but exhibit diverse trajectories. This led us to the use of an object tracking algorithm. To preserve the simplicity of the pre-processing stage, we utilized a straightforward centre of mass computation, involving only the summation of the synthetic frame's rows and columns (see fig.2a). By locating the centre of each particle, we cropped the dimensions of the synthetic frame to 100x100 pixels. Based on this pipeline, 4378 unique synthetic frames were captured/generated, not equally divided among the three classes: 1216 particles correspond to 20 μm spheres, 1811 to 16 μm, and 1351 to the 12 μm class. A critical issue in IFC data processing, as raised in [24], is related to whether experimental bias can influence the accuracy of ML schemes; specifically, if measurements of two particle classes are performed under different experimental conditions, then the ML's accuracy might reflect these experimental biases rather than the inherent differences between the two classes. To avoid this pitfall, we recorded data through multiple experimental instances (on different days, using different microfluidic channels-replaced due to clogging), ensuring multiple classes were recorded in each instance. In Fig. 3a-c, images from three typical microfluidic channels used in different instances are presented. It is evident that minor positioning errors among instances can affect the IFC magnification factor (width of channel in pixels). To address this, we normalized each synthetic frame's spatial features (measured in pixels) to the width of each microfluidic channel used. In Fig. 3d-f, normalized particle size histograms for the three classes are presented; each class exhibits a distinguishable mean value ($m_{d=20\mu m} = 0.25 \pm 0.04$, $m_{d=16\mu m} = 0.17 \pm 0.02$ and $m_{d=12\mu m} = 0.14 \pm 0.016$), while significant overlap between the three distributions is observed. This implies that even after data normalization, the classification task remains challenging, necessitating an ML model.

**Digital Machine Learning IFC**

In this direction, we focused primarily on FPGA-compatible and lightweight ML models. A subset of the dataset was randomly chosen to equalize the number of samples per class. Therefore, 1216 unique samples per class were used, of which 70% were used for the training of each model and 30% for testing. All ML models were implemented on the TensorFlow framework[25] by a graphics processing unit (GPU) alongside an Adam Optimizer[26] with a learning rate of 0.01 and categorical cross-entropy as the loss function. The model training process involved data processed in batches of 250 for a maximum of 500 epochs. To prevent overfitting, an early-stopping callback was implemented to monitor validation loss, maintaining a validation split of 10% during training (see Methods). The architectures used here include a fully connected network with 1 and 2 layers. In Table I, we present the accuracy and the number of trainable parameters for the lightest neural network configuration employed: a single layer consisting of 3 perceptron nodes, each equipped with a rectified linear unit (ReLU) activation function. The optimum performance without digital pre-processing emerged when the synthetic frames are directly used as input (97.5%), requiring 60,000 FLOPs. In an approach to boost performance, we also utilized a digital pre-processing technique tailored for image processing such as Histogram of Gradients (HoG)[27]. In this case, accuracy marginally increased to 98% while FLOPs were reduced to 26,000. Nonetheless, these benefits come with the subtle cost of additional digital pre-processing of the original data. In particular, HoG entails the computation of the magnitude and the angle of each pixel's contrast to its adjustment pixels row/column wise, thus in this case leads to a minimum of $10^5$ additional computations during pre-processing[27]. It is worth mentioning that each synthetic frames' intensity histograms (used to extract Fig. 3) was used as input to the same neural network. This approach can act as basic benchmark of accuracy/complexity. In this context, although histograms require an extremely low number of parameters (303), their performance tops at only 90%, confirming that the IFC classification is not a trivial task.

| Table I: Performance of a single layer feedforward neural network for raw events, histograms and HoG | | | |
|---|---|---|---|
| Metric | Best Accuracy (%) | Trainable Parameters | FLOPS |
| Synthetic Frames (100,100) | 97.5 | 30,003 | 60,001 |
| Histogram (100) | 90 | 303 | 601 |
| HoG (66,66) | 98 | 13,071 | 26,137 |

Aiming to further boost performance, we use a feedforward neural network with an additional dense layer. In Table II, we present performance metrics including accuracy, number of trainable parameters and power consumption, using the thermal design power (TDP) for each network. For each metric, we compute two values: one for the ML model with the fewest parameters (highlighted in bold) and another for the model providing the best performance in terms of accuracy (indicated in regular font). Using a compact 2-FCL network with only 6 hidden nodes achieves a performance of 97.9%. However, increasing the FCL parameters radically to 46 hidden nodes yields only a minor improvement of 0.3%, with an accuracy of 98.2%. Similar to the single FCL case, the highest accuracy is achieved through HoG pre-processing

(98.3%), which halves the number of trainable parameters but requires more complex data pre-processing, as mentioned above. The best overall performance, a 98.7% accuracy, is again achieved with HoG and 2 FCLs. Interestingly, the accuracy boost when using a two-layer FCL network over a single-layer perceptron is marginal. For example, the accuracy increases from 97.5% to 98.2% for synthetic frames used as direct input and is accompanied by a significant rise in FLOPs, from 30K to 900K.

**Table II:** Performance of a 2-FCL neural network. The metrics in bold correspond to the most lightweight network in terms of parameters and the metrics in regular font to the optimum network in term of performance. In the last column the number of nodes at the hidden layer and the type of non-linear activation function used are presented.

| Metric | Accuracy (%) | Trainable Parameters | FLOPS | Thermal Design Power % | Optimum nodes/function |
|---|---|---|---|---|---|
| **Synthetic Frames (100,100)** | **97.9**/98.2 | **60,027** / 460,187 | **120,038**/920,278 | **15.1**/19.4 | **6, relu** / 46, relu |
| **HoG (66,66)** | **98.3**/98.7 | **34,883** / 200,563 | **69,746** / 401,030 | **11.3**/16.9 | **8, tanh** / 46, tanh |

**Photonic Neuromorphic Pre-Processor for IFC**

a) **Numerical Simulations**

As shown in the results above, input pre-processing (e.g., HoG) is a potential path towards further increasing accuracy, without massively affecting the number of trainable parameters. On the other hand, digital pre-processing introduces additional computations, thus impacting power consumption during inference. Aiming to circumvent this impediment, we pursue an alternative approach, where input signals are pre-processed directly in the optical/analogue domain. In particular, we utilize a neuromorphic photonic scheme that will alleviate the additional computational workload associated with digital feature extraction techniques[28,29], while at the same time, off-loading part of the back-end computational complexity, allowing for slimmer digital back-end ML models. In this context, our group proposed a photonic accelerator, relying on a hardware-friendly optical spectrum slicing (OSS) technique[23]. This approach involves the utilization of multiple passive optical filters acting in parallel as convolutional neural nodes. Each OSS node monitors distinct spectral regions of the input optical signal and applies a complex kernel filter directly in the analogue domain without any need for digital processing. This scheme was successfully applied to the field of high-speed image processing, where the OSS has offered performance levels comparable to fully digital sophisticated architectures[23]. More importantly, OSS, when targeting datasets such as the MNIST, offered a significant reduction in the number of trainable parameters and consequently to the overall power consumption. The fact that recurrent versions of the OSS have been also tested in transmission impairment mitigation in high baud rate optical communications further solidifies the capabilities of this approach as a ML accelerator[30]. On the other hand, OSS has been only recently combined with real-life datasets and, in particular, with medical imaging modalities such as IFC[21]. Here, we aim to combine the efficiency of a neuromorphic camera with photonic neuromorphic pre-processing for the first time, so as to enhance the IFC's capabilities overall.

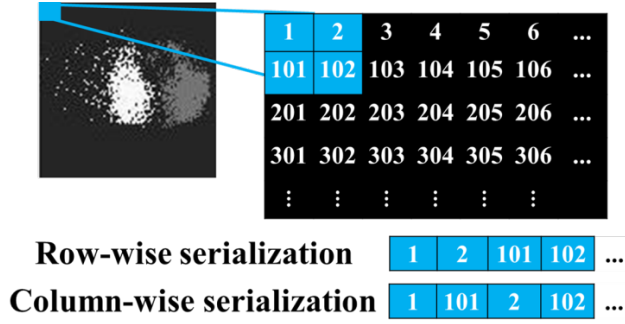

Figure 4: Conversion of synthetic frames to 1D vectors for OSS processing. Each frame is serialized into a 1D vector by processing blocks of pixel values (patches) in two orientations: row-wise and column-wise (see Methods).

Before implementing the OSS scheme, synthetic frames are converted into 1D vectors, as depicted in Figure 4. This conversion involves serializing each block of pixel values (patch) in two orientations—row-wise and column-wise. Both orientations are used in the 1D representation of each synthetic frame to enhance their spectro-temporal characteristics (see Methods section). The key components of the OSS scheme are illustrated in Figure 5. The 1D vectors of each synthetic frame are transferred to the optical domain by modulating the amplitude of a continuous-wave optical carrier through a digital-to-analog converter (DAC) and an electro-optic modulator, such as a Mach-Zehnder Modulator (MZM). The processing core comprises multiple bandpass optical filters. Operation wise, the application of the filter's transfer function in the frequency domain is identical to the convolution of the signal with the filter's impulse response in the temporal domain. Therefore, each OSS node is set at a different central frequency, facilitating the "slicing" of distinct regions of the optical signal. This results in a change in the impulse response of each filter and the application of varying complex weights to the input time-traces[23]. Unlike digital kernels, the control over the complex weights is coarsely adjusted through the filter's hyperparameters, which shape the impulse response of the OSS filters, such as central frequency ($f_m$), bandwidth ($f_c$) etc. (red and green insets of Figure 5). Furthermore, the filters can act as tunable optical integrators[31], where the integration time is governed through tuning the filters' bandwidth and order. The integration time in this scenario is equivalent to the receptive field; meaning the number of spatial pixels' values that are linked during convolution.

Following this purely optical step, the time-traces are transferred back to the electrical domain through a photodiode (PD) and an analog-to-digital converter (ADC) that follows each filter. The PDs introduce an elementwise nonlinear transformation at the filter outputs through their square-law characteristics. Furthermore, by reducing the bandwidth of the PDs or equivalently by integrating the PDs' output, an average-pooling-like operation is performed. The extent of this integration can be controlled through Eq. 2, where the PD's 3 dB bandwidth is set according to the pixel rate (modulation rate at the MZM) of the input signal (PR) and the size of the patch. This is defined by dividing the synthetic frame into square blocks of $M \times M$ pixels.

$$BW_{PD} = \frac{PR}{M \times M} \qquad (2)$$

Towards the same direction, the sampling rate (SR) of the ADCs can be used to further compress data by under-sampling the output. The digitized samples at the output of the OSS are subsequently flattened and fed to a simple digital back-end comprising a lightweight digital FCL with 1 or 2 dense layers, identical to the ones presented above. It is worth mentioning that the number of nodes in the FCL and thus the trainable parameters are directly governed by the number of samples that are generated and fed to the back-end.

(c)

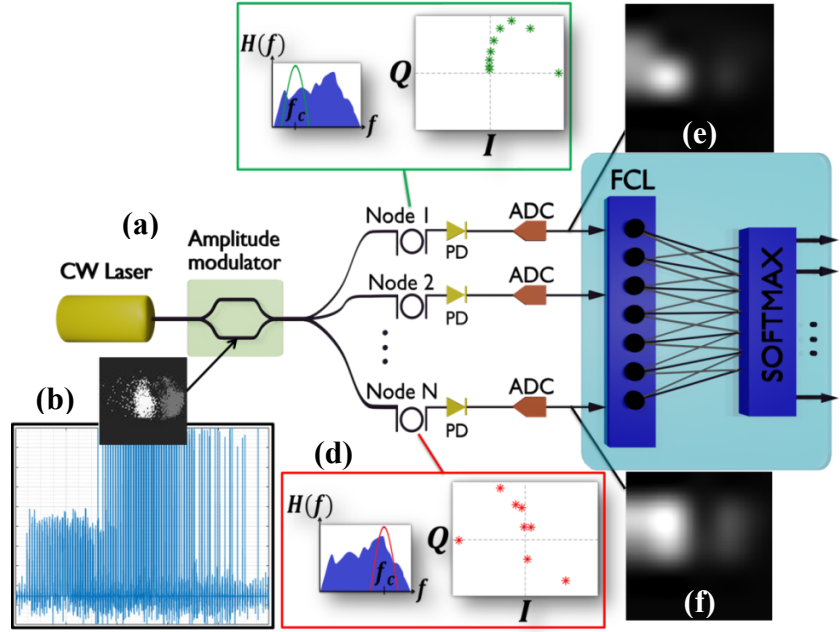

Figure 5: Schematic diagram of the OSS-CNN for the classification of IFC experimental data. (a) The synergy of a continuous-wave laser, a DAC and an MZM, used to imprint the 1D synthetic frame vectors onto the amplitude of the optical carrier. (b) A flattened IFC synthetic frame after laser modulation, displayed alongside the original synthetic frame. (c-d) Complex-valued coefficients (I/Q) for two identical filters, positioned at different detunings relative to the signal's carrier frequency. (e-f) Reconstructed images after processing through two discrete OSS nodes.

Towards validating the concept, the first step consists of numerically simulating the OSS scheme so as to determine the optimum hyperparameters that would maximize classification accuracy. In this case, the OSS nodes are implemented as first-order bandpass Butterworth filters that can be easily realized using conventional photonic structures (e.g micro-ring resonators-MRRs). The system's hyperparameters under examination included the patch size ($M \times M$), stride, number of OSS nodes and OSS filter characteristics ($f_c, f_m$). Initially, optical input was assumed to be injected to the filters at a pixel rate of 100 GS/s, with an average input power of 20 mW; a typical optical spectrum and time-trace of a particle are shown in Figures 6b and 6a, respectively. Subsequently, the optical signal was split equally to feed the OSS nodes with a 1xN splitter, where nodes range from 1 to 10. In Figure 6c, three typical filter responses are depicted at different frequencies, "slicing" the spectrum of the input signal. The bandwidth of the PDs was adjusted in accordance with the selected patch size, as defined in Eq.2, while the sampling rate of the ADCs was modified to extract only two samples over the temporal duration of a single patch.

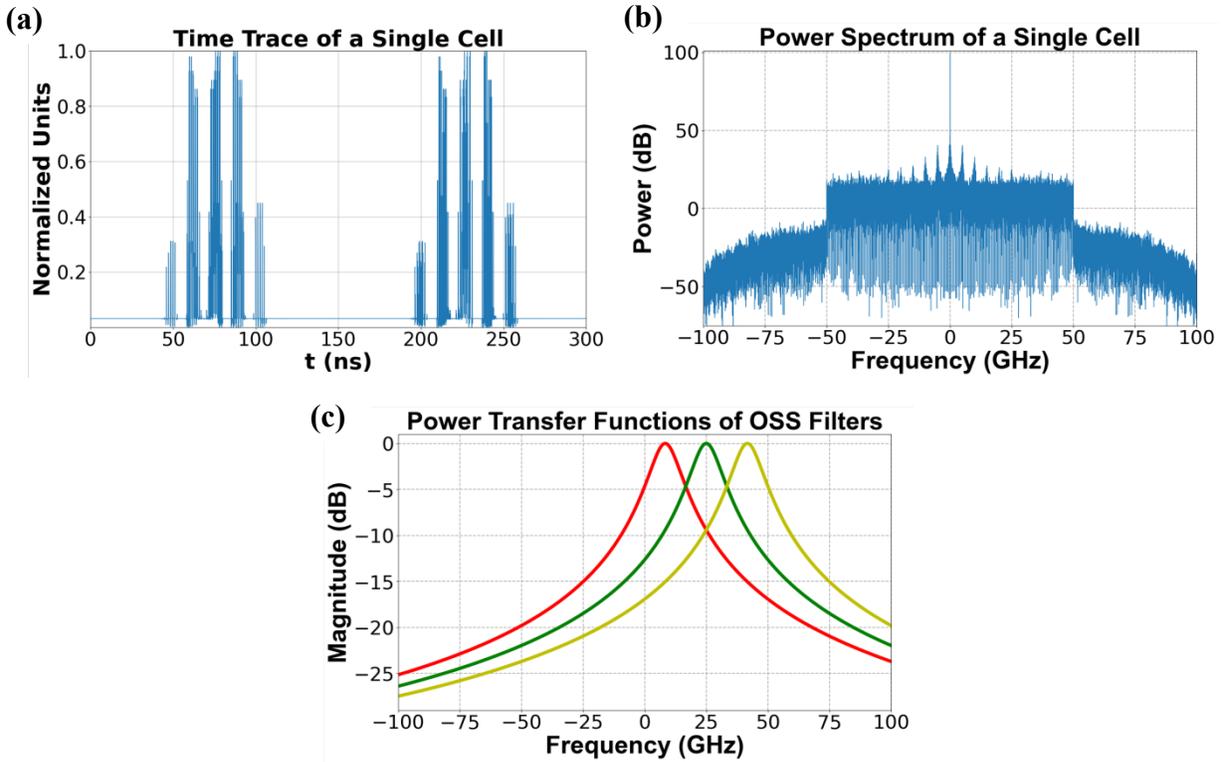

Figure 6: (a) Time trace at the output of the DAC corresponding to the 1-D vector of a 12 μm cell, (b) Optical power spectrum of the signal modulated by the vector of the 12 μm cell and (c) Power transfer functions of three OSS filters with a 12 GHz 3-dB bandwidth, designed for uniform segmentation of the right-sideband spectrum within the input optical signal.

The generated digital samples were fed to a single FCL identical to the digital scenario described above. The key differentiation in this approach lies in the preliminary optimization phase. Prior to classification, we employed the 'Optuna'[32] hyperparameter optimization framework to systematically identify the optimal values for the OSS scheme in this task (bandwidth, central frequency and patch size).

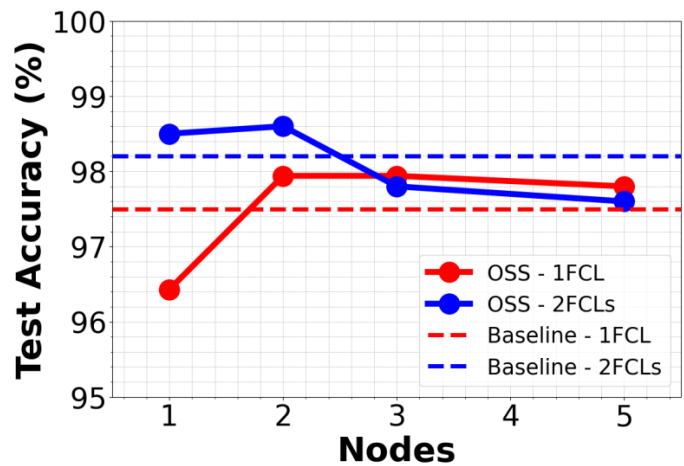

Figure 7: Classification accuracy of a numerically simulated OSS-CNN with one (red circles) and two FCLs (blue circles) across varying OSS nodes, compared to standalone one-layer (red dashed line) and two-layer FCL models (blue dashed line).

The hyperparameter scan revealed that using a 20x20 pixel patch with a 10-pixel stride, during the conversion of 2D image data to a 1D format for OSS processing, yielded optimal results. Figure 7 graphically illustrates the performance of two OSS-CNN configurations—comprising one and two FCLs—against the number of OSS nodes utilized. For context, the standalone performances of both single and dual-layer digital FCL models are included for direct comparison, depicted by red and blue dashed lines, respectively. In particular, utilizing a single OSS node, with a filter 3dB bandwidth ($f_m$) of 11 GHz and a central frequency detuning ($f_c$) of 19 GHz, testing accuracy reached as high as 96.4%. Adding just one more OSS node—adjusting $f_m$ to 16 and 37 GHz for each filter, with $f_c$ set to 8 GHz—enhanced classification accuracy to 97.9%, slightly surpassing the standalone single-layer FCL model's performance by 0.4% (see Table I). A similar accuracy level was also achieved in the cases that node number increased to 3-5. Further exploration with a two-layer FCL classifier, as indicated by the Optuna framework, included 80 nodes in the digital hidden layer with a ReLU activation. This setup showed a notable improvement, where the single-node OSS scheme's accuracy boosted to 98.5%, compared to 98.2% for the purely digital approach. The highest accuracy achieved with the OSS-assisted scheme was 98.6% with two optical nodes, though an increase in the number of OSS nodes beyond three resulted in a slight decline in performance, likely due to overfitting. Overfitting here arises when more than three OSS nodes are used, due to increase in the model's parameters.

From a first glance, the inclusion of a fully analogue accelerator in this case, does not provide a strong performance boost as in [23] but still improves testing accuracy by 0.5%. However, the key advantage of integrating OSS lies in its impact on reducing the number of trainable parameters, a critical factor in minimizing power consumption during model training. Table III provides a detailed comparison of OSS configurations (with one and two FCL layers) against purely digital neural network models, including the standalone single-layer and double-layer feedforward neural networks described above and a sophisticated yet parameter-efficient gated recurrent unit RNN (GRU-RNN). It highlights the differences in classification accuracy, the number of trainable parameters and the number of hidden units for each architecture.

| Table III: Comparison of Different Digital Neural Network Architectures of the Event-Based Data | | | |
|---|---|---|---|
| **Architectures** | **Hidden units** | **Parameters** | **Accuracy** |
| **1 FCL** | - | 30003 | 97.5% |
| **2 FCLs** | 46 | 460187 | 98.2% |
| **OSS-1FCL** | - | 1947 | 97.9% |
| **OSS-2FCL** | 80 | 52163 | 98.6% |
| **GRU-RNN** | 38 | 1144677 | 98.6% |

The results of Table III underscore that the OSS was able to achieve a classification precision higher compared to simple FCLs and identical to the more complex GRU-RNN architectures. This is particularly noteworthy since the OSS-2FCL configuration operates with significantly fewer trainable parameters, approximately 8, 15 times lower compared to 1-FCL and 2-FCL respectively, whereas versus the complex RNN, OSS offers a strong parameter reduction by a factor of 22, without any performance degradation. This key property, similar to [23], can be attributed to the CNN-like architecture of the OSS system. Specifically, the slicing process by multiple filters performs a convolutional operation, where the kernel is roughly dictated by the transfer function and position of each filter's response. This process extracts multiple diverse features from the original data, which are also integrated by the filter's integrational property. Additionally, these rich, information-wise features undergo a nonlinear transformation through the PD, which is also set to average these values over the temporal duration of a patch (see Eq.2). Consequently, the number of samples delivered to the input of the back-end classifier for each synthetic frame is significantly reduced, corresponding to two samples per patch as determined by the sampling rate of the post-nodal ADCs.

**b) Experimental Validation of OSS assisted IFC**

The aforementioned concept was experimentally validated, as depicted in the schematic diagram of Figure 8. The setup included a tunable laser (CoBrite-DX Tunable Laser) amplitude-modulated through a 20 GS/s arbitrary waveform generator (AWG) (Tektronix AWG70002B) and a Mach-Zehnder modulator (MZM) (iXblue MXIQ-LN-30) with a 20 GHz 3-dB bandwidth. Subsequently, the modulated signal was directed to an optical waveshaper (Coherent WaveShaper-1000A Programmable Optical Filter), acting as a Butterworth filter with a 3-dB bandwidth around 10 GHz, centred at 1552.6 nm, realizing a single OSS node. The optical output was captured by a 10 GHz photoreceiver (Thorlabs RXM10AF) connected to a 50 GS/s digital signal oscilloscope (OSC) (Tektronix DPO75002SX). The final digital signal was resampled to 20GS/s to match the rate of the AWG and was filtered using a digital low-pass filter to implement an operation akin to average pooling. The final step involves classification using a digital dense feedforward neural network, identical to the one used in the numerical simulations described above.

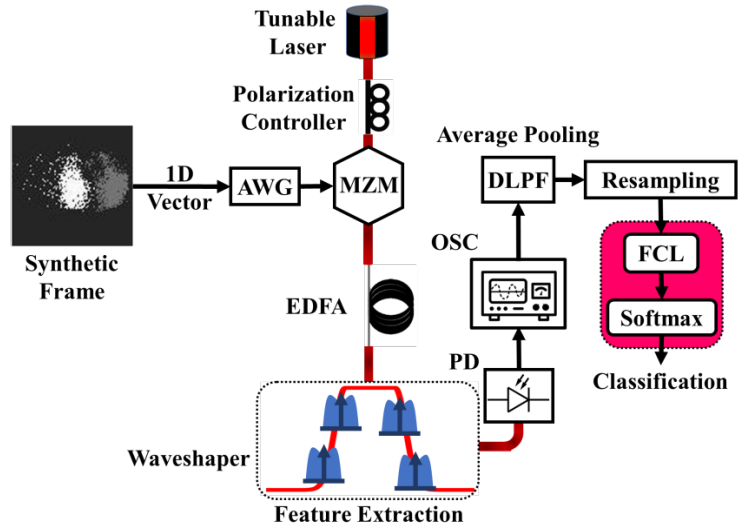

Figure 8: Schematic diagram of the OSS-CNN experimental setup. The setup includes a tunable laser, a 20 GS/s AWG, and an MZM. The signal is then boosted by an EDFA to enhance SNR and sent to a waveshaper. Detection and analysis are conducted with a 50 GS/s OSC. Post-processing is performed using a DLPF and classification via a FCL.

The data pre-processing pipeline was slightly modified compared to the numerical simulations, aiming to reduce the number of samples processed by the AWG due to time constraints. In this approach, we took the original 100x100 pixel frames and divided them into smaller, 4x4 pixel blocks. By calculating the average value of each 4x4 block, we effectively compressed the frame into a smaller, 25x25 pixel frame, reducing the overall data. Therefore, the compressed frames were then transformed into 1D vectors using 5x5 patches and a stride of 2. To increase the SNR of the optical signal and assess the system's capabilities when not limited by thermal/shot noise during detection, an Erbium-Doped Fiber Amplifier (EDFA) was used prior to the waveshaper to boost the optical power to 8 dBm. To simplify the experimental setup and facilitate the recording of a single trace at a time, the optical convolution process at the waveshaper was performed sequentially. Specifically, the filter response of the waveshaper was maintained constant at 193.22 THz (1552.6 nm), while to implement multiple OSS nodes, the central frequency of the laser source was detuned relative to the central frequency of the filter. The carrier frequency was adjusted in 1 GHz steps, ranging from -9 GHz to +9 GHz, allowing for the exploration of 19 distinct filter central frequencies. Figure 9a shows a segment of a normalized time trace corresponding to a 12 μm particle as sent to the AWG. In Figure 9b, the resulting outputs for two different filter-carrier detunings (-6 and +1 GHz), as captured by the oscilloscope, are presented. The variation in frequency detuning led to distinct outputs, highlighting the optical filter's differential interaction with the signal's spectral components.

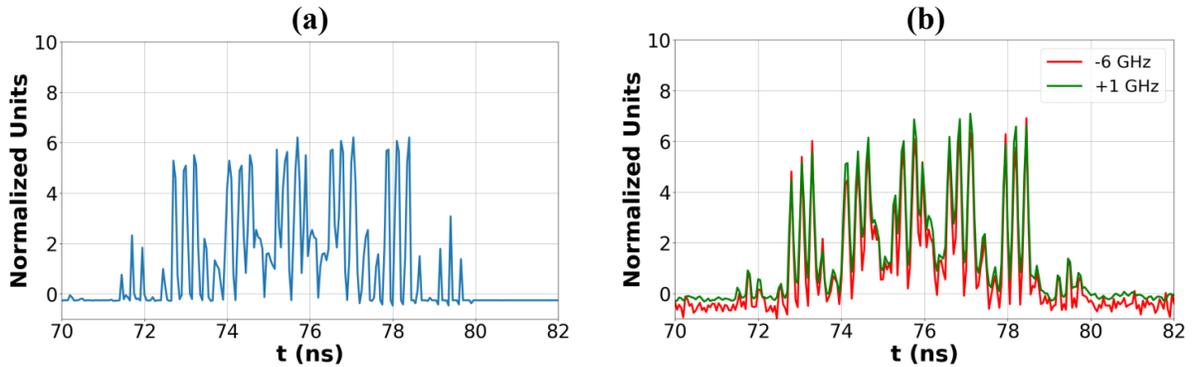

Figure 9: (a) Temporal segment of the normalized event-vector time trace from the AWG corresponding to a 12 μm cell. (b) Normalized outputs for two distinct filter detunings from the carrier frequency at the OSC, specifically at -6 GHz (red) and +1 GHz (green), applied to the input signal depicted in (a).

After photodetection, the signal from the oscilloscope was resampled at a rate of 20 GS/s to align with the AWG's output rate. The number of consecutive samples subjected to averaging was then adjusted based on the bandwidth of the following LPF, as per Eq.2, leading to varying degrees of compression at the output of the OSS system. The compression ratio (CR) serves as a metric that demonstrates the level of compression, compared to the input, applied at the OSS accelerator's output. It is defined as the ratio of the initial data size to the size of the digital outputs of the OSS nodes per dataset sample. For this experimental setup, the compression ratio can be calculated using the formula:

$$CR = \frac{N}{2N_f} \qquad (3)$$

where $N_f$ is the number of distinct filter positions contributing to the inputs for the FCL and $N$ is the integer denoting the number of successive samples involved in the averaging process. The averaged outputs from all particle samples were then merged to construct the optical pre-processed version of the original dataset.

The classification process began with analyzing the digitized outputs from each filter position, thereby examining the performance when a single OSS node was used for pre-processing under various averaging scenarios. Subsequently, the back-end FCL was supplied with combinations of outputs from 2, 3, and 5 different OSS nodes, thereby implementing an expanded OSS scheme. Figure 10 illustrates the mean classification accuracy for the optimal combination of two distinct filter positions (optimum found to be -1 and 0 GHz from the carrier) with a single-layer FCL as the digital classifier. This is compared alongside the performance of the optimal simulated 2-node OSS (solid-blue line), in relation to the CR. Additionally, Fig. 10 presents the accuracy of the standalone digital FCL processing the entire synthetic frame dataset (black dashed line) and the accuracy from a time trace derived from an uncompressed synthetic frame without any OSS node intervention (red dashed line). Here, "mean accuracy" refers to the average classification accuracy obtained over ten iterations of the same 3-neuron FCL model, where the only difference across iterations was the initial values of the weights, with all the other training parameters remaining constant. From Fig. 10, it is evident that the performance of the experimental OSS setup is similar to that of the simulated OSS. The highest accuracy recorded in the experimental setup was 98.1% at a CR of 3.3, underscoring the efficacy of OSS by enhancing accuracy by 0.6% over a system without noise, while concurrently reducing the number of trainable parameters by a factor of 3.3. Another significant observation from Fig. 10 is the robustness of OSS pre-processing, which sustains nearly 96% accuracy even under substantial compression (96.1% with a CR of 20).

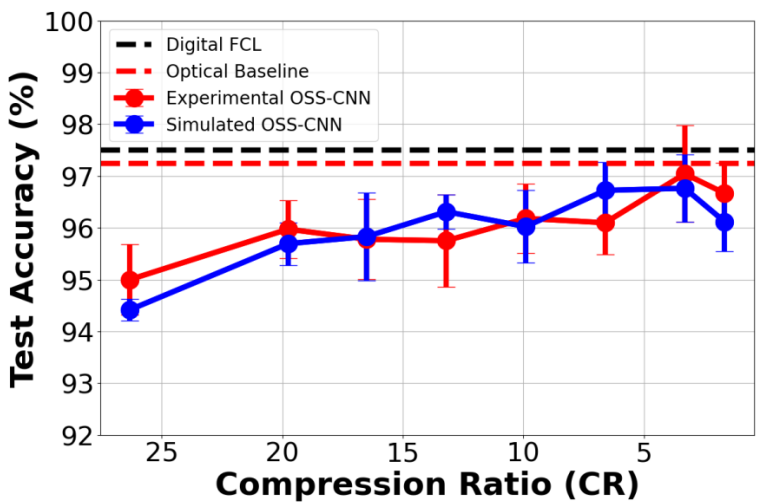

Figure 10: Mean classification accuracy for the optimal 2-node experimental OSS configuration, utilizing filters detuned by -1 and 0 GHz, paired with a 1-FCL classifier, in comparison to a simulated 2-node OSS as a function of the CR. For reference, the accuracy of a standalone FCL is shown (black dashed lines), alongside the mean accuracy from the uncompressed output of a wideband (25 GHz) filter aligned with the carrier frequency (red dashed lines) are also presented.

In Figure 11, the mean precision of the experimental OSS-CNN is plotted against the number of OSS nodes (filter positions), considering a single-layer FCL (Fig. 11a) and a dual-layer FCL backend (Fig. 11b), with the CR held constant at 3.3. The figure also compares the classification accuracy of the simulated lossless OSS at the same CR and the accuracy achieved by digital single-layer and dual-layer FCLs without OSS pre-processing. From Fig. 11a, the accuracy trend of the experimental setup is shown to be similar to the simulated OSS, demonstrating comparable or higher accuracy than the standalone digital FCL when more than two OSS nodes are used. Notably, the highest accuracy achieved experimentally is 98.1% with three nodes, surpassing the single-layer digital FCL despite the experimental system being subject to noise (photodetection, EDFA etc.). The experimental OSS system's robustness is evident in its ability to maintain a precision level above 97% with 1, 2 and 3 nodes. Fig. 11b reveals that incorporating a second dense layer in the digital backend enhances accuracy across all cases compared to Fig. 11a. The peak accuracy for the experimental OSS was 98.4% with 1-2 OSS nodes, marginally lower than the 98.6% peak performance of the simulated OSS. The experimental setup's overall performance slightly declines across node configurations, highlighting the influence of additional experimental parameters, most notably in the 2-node case where mean accuracy dips to 97.6%, compared to 98.6% in the simulated system. Interestingly, in the 1-2 OSS nodes configurations, the optimal experimental OSS surpasses the dual-layer digital FCL while also significantly reducing the number of trainable parameters.

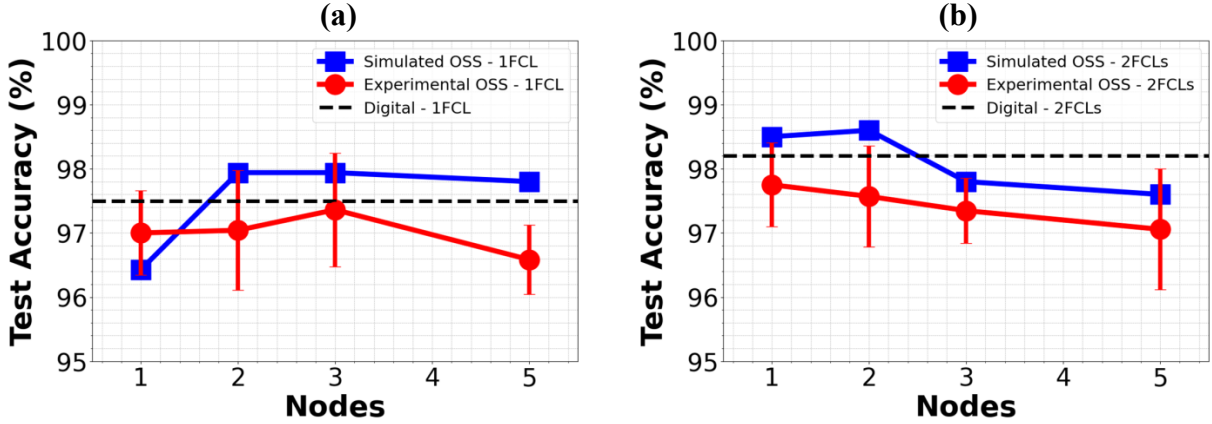

Figure 11: Mean classification accuracy of the experimental OSS versus the number of nodes used with: (a) a single-layer (1-FCL) and (b) a two-layer (2-FCL) digital back end. As a reference, the classification accuracies of the numerically simulated lossless OSS and the maximum accuracies achieved by the standalone noiseless FCL with one and two layers are included for comparison.

**Discussion and Conclusion**

In this work, experimental data from a neuromorphic event based IFC system have been recorded, regarding the classification of aqua solution of PMMA particles with different diameter. The generated datasets comprise rate-encoded synthetic frames subject to a lightweight pre-processing, mainly consisting of object tracking, to compensate for the utilization of a generic micro-fluidic system. Classification of the IFC data has been performed in three ways: lightweight digital feedforward neural networks, offering a classification accuracy of 98.2% comparable to [2], but requiring more than 460K trainable parameters. The second approach consisted of using digital pre-processing of the dataset, following the well-established HoG algorithm; offering topping accuracy at 98.7% with 200K parameters during training and with the additional digital processing cost during inference. The third route replaces the HoG algorithm with an all-optical convolution-like photonic neuromorphic scheme, based on optical spectrum slicing; this route offloads the pre-processing to the analogue domain while the same feed-forward neural networks are used as back-end. In this work this concept has been validated both experimentally and numerically and it offered an optimum accuracy of 98.6%. The most critical attribute of the proposed scheme is that this performance is achieved with only a fraction of trainable parameters demanded by standalone digital neural networks and without any type of computationally intensive digital preprocessing. Thus, reducing power consumption both during inference and training. These improvements stem from the ability of the OSS to inherently correlate pixel information and apply random, yet complex kernels directly in the analogue domain. This in turn, evokes a coarse projection of image's features to a higher dimensional space, like a reservoir computing scheme, thus simplifies classification by a lightweight digital neural network. On the other hand, the proposed IFC entails an additional electro-optic conversion step that is not present in standalone digital solutions. This conversion step is a necessity in all photonic accelerators[33,34], whereas for the explicitly OSS it has been shown that the power consumption of the electro-optics is lower compared to the energy-footprint of the digital back end. In addition, this step can be omitted if an inherently photonic IFC detector is employed as in [12]. Overall, the results in this work confirm that by matching neuromorphic sensing with neuromorphic processing an overall performance enhancement can be achieved outperforming all previous schemes, whereas offering a strong reduction in terms of trainable parameters by a factor >20 that is of utmost importance for emerging machine learning modalities.

# Methods

## Experimental Setup

The experimental setup features two 40x microscope objective (Marka) lenses with a numerical aperture (NA) of 0.65 and a working distance of 600μm. These lenses are securely mounted on two 3-Axis MicroBlock. Two planoconvex lenses with a focal length of 10cm are used before and after the objectives so as to direct/collect light from the microscope system. The light source is a 635nm emitting LED with average power of 5mW, whereas the recording event-based camera is Gen4 provided by Prophesee with a resolution of 640x480 pixels and a temporal resolution in spike generation of 1μsec. The camera records pixel's contract changes with two polarity values, depending on whether there is an increase or decrease in intensity. In this work, synthetic frames do not include this feature and all events are mapped to the synthetic frame, independently of their polarity. A vacuum pump installed offered a steady liquid flow ranging from 10 μlit/hour to several ml/min thus, regulating the speed of the particles from 0.001m/sec to well beyond 1m/sec. In the experiments a particle speed of 0.07 m/sec was used. Finally, the microfluidic channels employed, were straight channels based on TOPAS, offering absorption below <0.5dB at 635nm, having a cross section of 100x100μm. The microfluidic scheme had no seethe control, thus particles propagated at random trajectories within the cross-section.

## Sample Preparation

The experiments involved three distinct categories of calibrated transparent PMMA spheres (POLYAN) of different diameters 12, 16, and 20 μm. The initial concentration of calibrated transparent PMMA spheres was 5% and it was further diluted to 1:200 by adding purified water. The low particle concentration was used so as to avoid clogging in the microfluidic channel and reduce the probability of particle clustering. All tubes and microfluidic chambers employed during dataset generation were used solely for one class to avoid cross-contamination that would hinder labeling of the dataset. Prior to any measurement, all the microfluidic channels were thoroughly cleaned by debris by injecting purified water at high flow.

## Synthetic Frames to 1-D Vectors

In order to leverage the OSS optical preprocessing, synthetic frames are serialized using multiple spatial orientations. Although this step does not require any digital preprocessing, it determines the order in which pixels are inserted into the stream, thereby potentially augmenting input data. Specifically, it can generate additional correlations among pixels when combined with a filter. The process consists of the following steps: the values of the synthetic frame are divided into blocks of size $MxM$ pixels, where $M$ is equal to 20 in numerically simulated OSS and to 5 in experimental OSS (see Fig. 4). Each block is sequentially serialized in two alignments: column-wise and row-wise. Both of these serialized forms are utilized in the representation of each synthetic frame in a one-dimensional vector. The stride, defined as the step by which the patch window shifts across the synthetic frame before serializing the next block, is employed in two orientations. The first orientation is row-wise, serializing pixels within the patch based on horizontal alignment, and the second is column-wise, focusing on vertical alignment of the patch pixels. Utilizing both patch orientations aims to augment the spatio-temporal characteristics of each particle sample within its one-dimensional representation. Consequently, the classification dataset comprises 4378 different one-dimensional vectors, each representing a synthetic frame.

## Machine Learning Training Process

The neural network models were formulated using the Keras API[35] implemented on the Tensorflow framework[25], and their training and evaluation processes were conducted utilizing a graphical processing unit (GPU). In all instances, 70% of the available data particles were designated for training, while 20% out of these were reserved for validation to mitigate the risk of overfitting. The remaining 30% of the data were allocated for evaluation to ensure a comprehensive and unbiased assessment of model performance. The optimization algorithm selected for these models was Adam, with a fixed learning rate of $1 \times 10^{-4}$ and the training regimen was performed for 300 epochs. The architectural configurations encompassed a basic digital perceptron, a fully connected network (FC) with 1-2 layers and lightweight recurrent neural networks (RNN), instantiated as long-short term memory (LSTM), gated-recurrent unit (GRU) and Vanilla-RNN (V-RNN).

To optimize hyperparameters, the 'Optuna' framework[32], which utilizes a Tree-structured Parzen Estimator, was employed to maximize the testing accuracy. A comprehensive evaluation of the aforementioned models was undertaken, focusing on several factors: the number of hidden neurons, the learning rate and the batch size within the FCs or the RNN models, as well as the activation function of the hidden nodes in the FC layers.

The data delineated in Tables II and III encapsulate a robust array of computational characteristics, including the number of parameters, FLOPS, and Thermal Design Power (TDP). The TDP represents the maximum amount of heat a processing component, like a GPU is expected to generate under heavy loads, where a lower TDP generally indicates reduced power consumption. These metrics were derived through the application of TensorFlow libraries in conjunction with NVIDIA tools, ensuring precise and dependable measurements critical for assessing and contrasting computational performance. In this analysis, an NVIDIA 2080Ti GPU with a TDP of 250 Watts was utilized. The determination of the TDP percentage was conducted using the GPU-Z monitoring utility[36]. The average TDU percentage was calculated over 100 training epochs for each neural network model.

Data availability

*All codes generated for the emulation of the optical spectrum slicing concept are uploaded in repositories alongside the processed synthetic frames from the cytometry experiments.*

Competing interests (mandatory)

The authors declare no competing interests.

Acknowledgements

This work has received funding from the EU H2020 project NEoteRIC under grant agreement 871330 and by EU Horizon PROMETHEUS project under grant agreement 101070195.

Author contributions

I. T and C. M setup and conducted the cytometry experiments, C. M did the digital pre-processing on the dataset, A. T performed the photonic pre-processor simulations and conducted the photonic machine learning experiment with help from G.S and A.B. S. D analysed the data using digital machine learning for benchmark reasons, K. P and G. B provided know-how and specifications on the event-based camera, A.B and C.M initiated the activity and supervised the whole process. The manuscript has been prepared with the help of all authors.